\pdfoutput=1 
\documentclass[a4paper]{jpconf}
\usepackage{graphicx}
\usepackage[numbers,square,sort&compress]{natbib}  
\usepackage{iopams}
\usepackage{float}
\usepackage{amsmath}
\usepackage{bm}
\usepackage{cleveref}

\begin{document}
\title{Predicting the benefit of wake steering on the annual energy production of a wind farm using large eddy simulations and Gaussian process regression}

\author{Daan van der Hoek$^{\mathsf{1}}$, Bart Doekemeijer$^{\mathsf{1}}$, Leif Erik Andersson$^{\mathsf{2}}$ and Jan-Willem van Wingerden$^{\mathsf{1}}$}

\address{$^{\mathsf{1}}$Delft University of Technology, Delft Center for Systems and Control, 2628 CD Delft, The Netherlands}
\address{$^{\mathsf{2}}$Department of Engineering Cybernetics, Norwegian University of Science and Technology, 7491 Trondheim, Norway}

\ead{ \{d.c.vanderhoek, b.m.doekemeijer, j.w.vanwingerden\}@tudelft.nl, leif.e.andersson@ntnu.no}

\begin{abstract}
In recent years, wake steering has been established as a promising method to increase the energy yield of a wind farm. Current practice in estimating the benefit of wake steering on the annual energy production (AEP) consists of evaluating the wind farm with simplified surrogate models, casting a large uncertainty on the estimated benefit. This paper presents a framework for determining the benefit of wake steering on the AEP, incorporating simulation results from a surrogate model and large eddy simulations in order to reduce the uncertainty. Furthermore, a time-varying wind direction is considered for a better representation of the ambient conditions at the real wind farm site. Gaussian process regression is used to combine the two data sets into a single improved model of the energy gain. This model estimates a 0.60\% gain in AEP for the considered wind farm, which is a 76\% increase compared to the estimate of the surrogate model.
\end{abstract}

\section{Introduction}
Currently, research on wind farm control mainly focuses on improving the energy yield of a wind farm. One promising method to achieve this is wake steering, in which upstream turbines are misaligned with the inflowing wind in order to redirect their wake away from downstream machines \cite{Jimenez2009}. Consequently, the downstream turbines experience a higher wind speed and produce more power, at the cost of a small reduction in power at the upstream turbines. This leads to a net increase in power production of the farm. The optimal yaw misalignments required to steer the wake are generally obtained using a simplified surrogate model of the wind \mbox{farm \cite{Boersma2017b}}. In order to validate the hypothesized energy gains from surrogate models, the control solutions are either implemented in high-fidelity simulations \cite{Gebraad2016}, wind tunnel experiments \cite{Campagnolo2016b} or field \mbox{tests \cite{Fleming2017,Howland2019}}. 

Field tests are expensive to perform, carry a certain risk and are difficult to validate. Furthermore, wind tunnel experiments only allow limited validation due to the scale and conditions that can be replicated. Simulations offer a cheaper and more practical solution and allow better control of the tested conditions, but they suffer from being very computationally expensive. For these reasons, wake steering is generally only validated for a small number of cases, and therefore does not provide an estimate of the potential increase in Annual Energy Production (AEP) for a wind farm. Instead, such predictions often rely on the surrogate model that was used for the yaw setpoint optimization in combination with statistical data obtained from measurements at the wind farm site \cite{Fleming2016journal,Gebraad2017,Kanev2018}. While these approaches predict energy gains in the order of 1\% for specific wind parks, the relatively low fidelity of such a surrogate model compared to reality casts a large uncertainty on these gains

Besides model mismatches that can influence the estimated AEP gain, the aforementioned approach does not take a dynamic implementation of wake steering and its additional challenges into account. These challenges include the estimation of ambient conditions for determining the optimal yaw angle, a realistic yaw setpoint tracking controller, evaluating the effects of wake steering on turbine loads, and robustness to time-varying ambient conditions. While often unaddressed, these aspects can have a major impact on the AEP benefit of wake steering. 

This paper evaluates the benefit of wake steering on the AEP of a large offshore wind farm using high-fidelity simulations and Gaussian process regression \cite{rasmussen2006gaussian}. It extends the work presented in \cite{Churchfield2015}, where several large eddy simulation (LES) results of a simple wind park were used to fit a curve of the expected energy gain. The main contribution of this paper is a framework that combines both low- and high-fidelity simulations into a single model of the energy gain, as a function of wind direction and wind speed. A surrogate wake model is used to predict the general trends for wake steering, while high-fidelity simulations are used to obtain a quantitative evaluation of wake steering for specific conditions. By combining these two approaches, a more realistic estimate of the potential gain for all ambient conditions is obtained, while running only a subset of high-fidelity simulations. 

The outline of this article is as follows. \Cref{sec:case_study} introduces the wind farm that is the main subject of this case study, as well as the simulation environments that will be used. In \Cref{sec:methodology}, the framework for estimating the annual energy gain will be presented. Simulation results of both low and high-fidelity models, along with the final estimate on the AEP are presented in \Cref{sec:results}. Finally, the results are discussed and the article is concluded in \Cref{sec:conclusion}.

\section{Case study} \label{sec:case_study}
The case study in this work is performed with the Princess Amalia Wind Park. This wind park has been the topic of previous studies on the effect of wake steering on the annual energy production, where different steady-state wake models predicted an increase in the AEP between 1.10-1.28\% \cite{Fleming2016journal,Kanev2018}. These studies will serve as a benchmark to which we compare our results later on. 

\subsection{Princess Amalia Wind Park}
\begin{figure}[t]
    \centering
    \includegraphics[clip,trim=6.5cm 11.cm 6.5cm 8.3cm,width=0.4\textwidth]{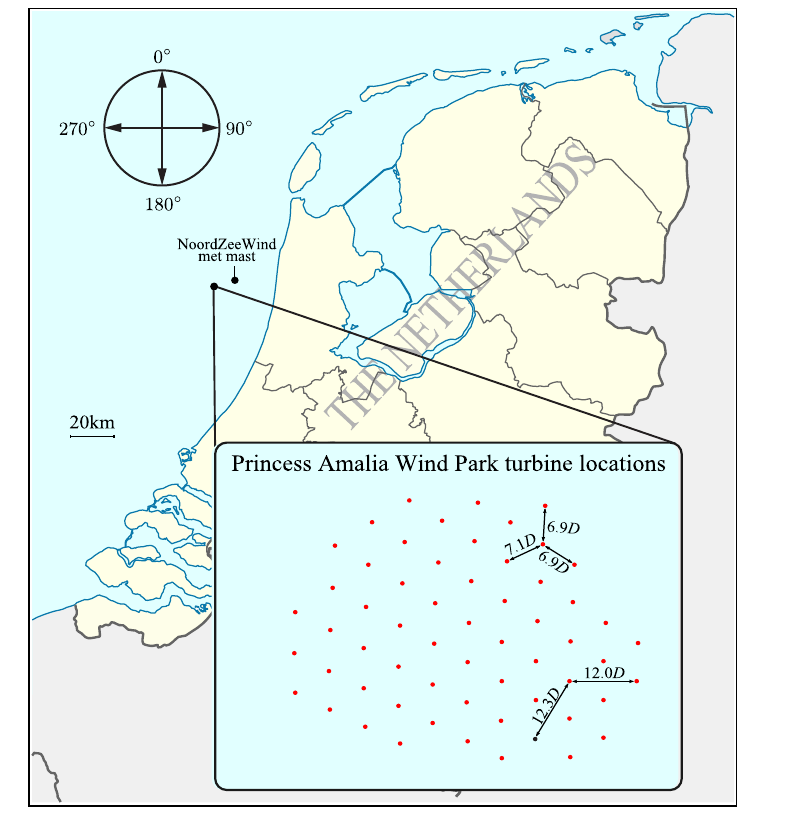}
    \caption{\label{fig:layout} Location and layout of the Princess Amalia wind park \cite{Gebraad2015b}.}
\end{figure}

The Princess Amalia wind park is located near the Western coast of the Netherlands and consists of 60 Vestas V80-2.0 MW wind turbines. The layout of the wind park is presented in \Cref{fig:layout}. Since the model and controller of the original Vestas V80 wind turbine are not publicly available, the turbines are replaced with the NREL 5MW wind turbine model \cite{Jonkman2009}. In order to make a fair estimate of the potential increase in energy for the real wind farm, the layout is scaled according to the dimensions of the NREL 5MW turbine. 

\begin{figure}[b]
    \begin{minipage}{0.49\textwidth}
        \centering
        \includegraphics[width=0.9\textwidth]{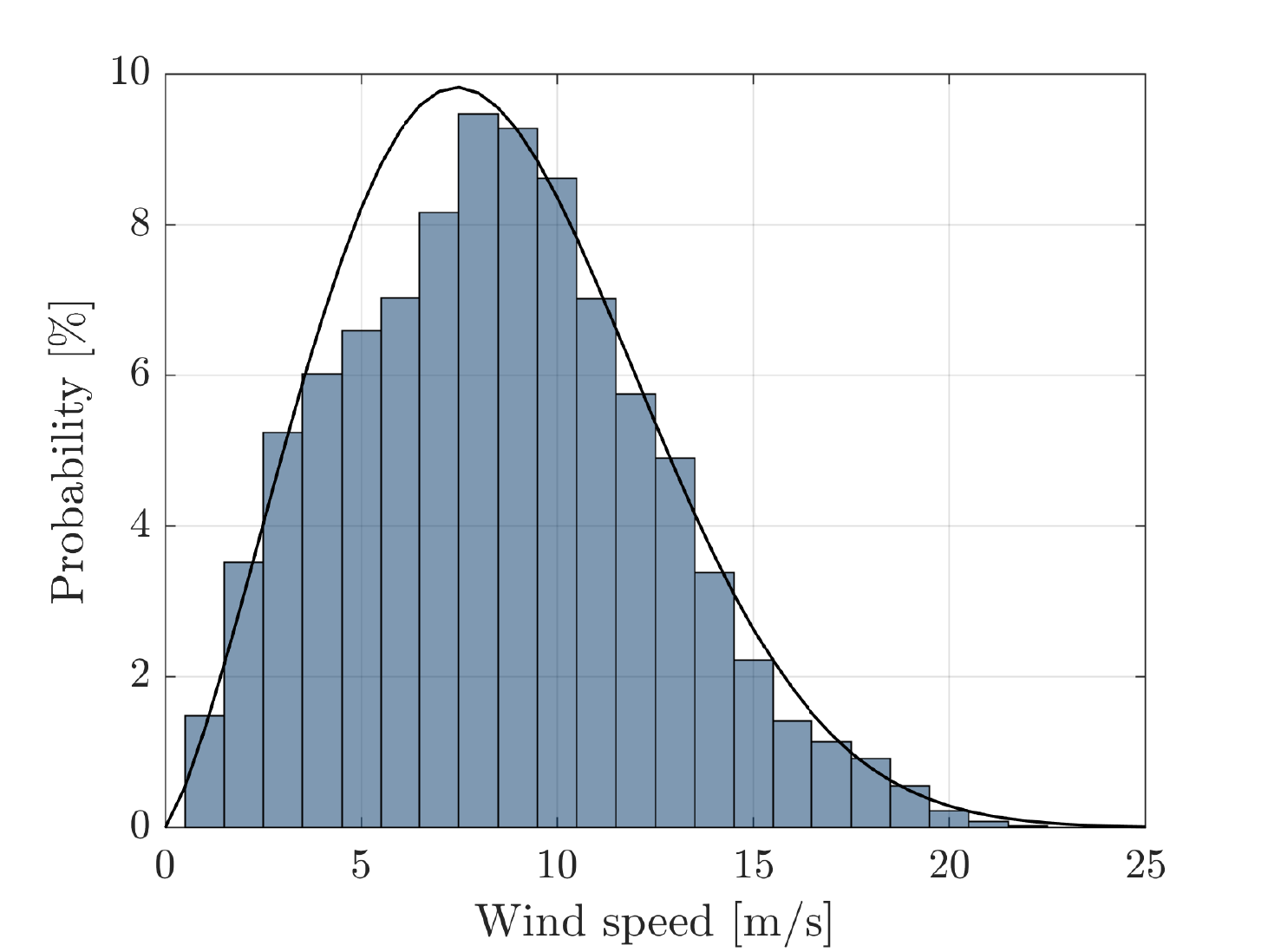}
        \caption{Wind speed distribution with accompanying Weibull fit from measurements collected at the wind farm site.}
        \label{fig:weibull}
    \end{minipage}\hspace{0.02\textwidth}%
    \begin{minipage}{0.49\textwidth}
        \includegraphics[width=0.9\textwidth]{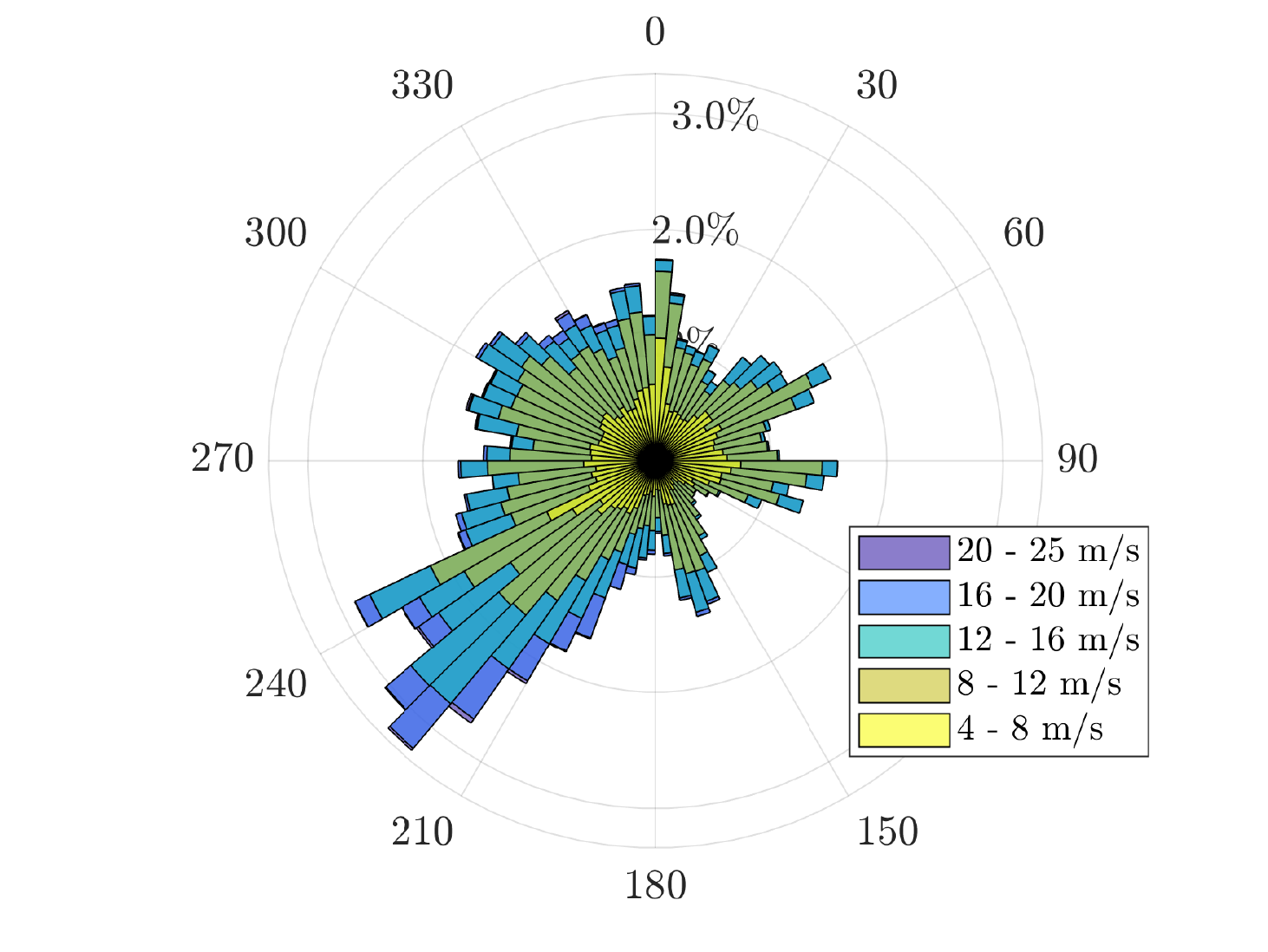}
        \caption{Wind rose from measurements collected at the wind farm site.}
        \label{fig:windRose}
        \vspace{4.2mm}
    \end{minipage} 
\end{figure}

Measurements taken from a nearby met mast for a period period of one year prior to the construction of the wind farm were used to determine the wind speed and wind direction distributions at the wind farm site \cite{Noordzeewind}, as seen in \Cref{fig:weibull,fig:windRose}. The average turbulence intensity at below rated wind speeds is close to $6\%$. The turbulence intensity is assumed to be constant for all wind speeds during the simulations. Additionally, the standard deviation of the wind direction was found to be approximately $3^{\circ}$. 

\subsection{Wind farm models}
The surrogate model FLOw Redirection and Induction in Steady-state (FLORIS) is used to optimize the yaw angles for below rated ambient conditions \cite{FLORIS2018}. For a given inflow wind field, FLORIS computes the time-averaged flow field and turbine performance inside a wind farm as a function of the turbine control settings. Parameters of the FLORIS model were tuned to high-fidelity simulation data in previous work \cite{Doekemeijer2019}. For more information on FLORIS, the reader is referred to \cite{Gebraad2016,Doekemeijer2019}. 

The optimal yaw setpoints that are obtained using FLORIS are evaluated with the Simulator fOr Wind Farm Applications (SOWFA), which is an LES model for wind farms developed at NREL \cite{Churchfield2012,sowfa}. The flow is simulated by solving the three-dimensional unsteady Navier-Stokes equations over a discretized domain. The wind turbines are modeled in SOWFA using a rotating actuator disk model (ADM-R) \cite{Wu2011}. The original wind farm layout is scaled to match its properties to the NREL 5MW turbine. An overview of the simulation settings is given in \Cref{tab:SOWFAsetup}.

\begin{table}[h]
    \caption{SOWFA simulations settings.}
    \label{tab:SOWFAsetup}
    \footnotesize\rm
    \centering
    \begin{tabular}{@{}*{7}{l}}
    \br
    Parameter & Value\\
    \mr
    Timestep & 0.5 s\\
    Simulation length & 3000 s\\
    Atmospheric stability & Neutral\\
    Domain size  & 9.0 km x 9.0 km x 1.0 km \\
    Cell size (outer region) & 20.0 m x 20.0 m x 20.0 m\\
    Cell size (up to 240.0 m altitude) & 10.0 m x 10.0 m x 10.0 m\\
    Blade epsilon & 20.0 m\\
    Free stream wind speed $U_{\infty}$ & 6.0, 8.0 and 10.0 m/s\\
    Free stream turbulence intensity $I_{\infty}$ & 5\%\\
    Free stream wind direction (average) $\phi$ & 240.0$^{\circ}$\\
    \br
    \end{tabular}
\end{table}

\subsection{Time-varying wind direction}
\begin{figure}[h]
    \begin{minipage}{0.49\textwidth}
    \includegraphics[width=1\textwidth]{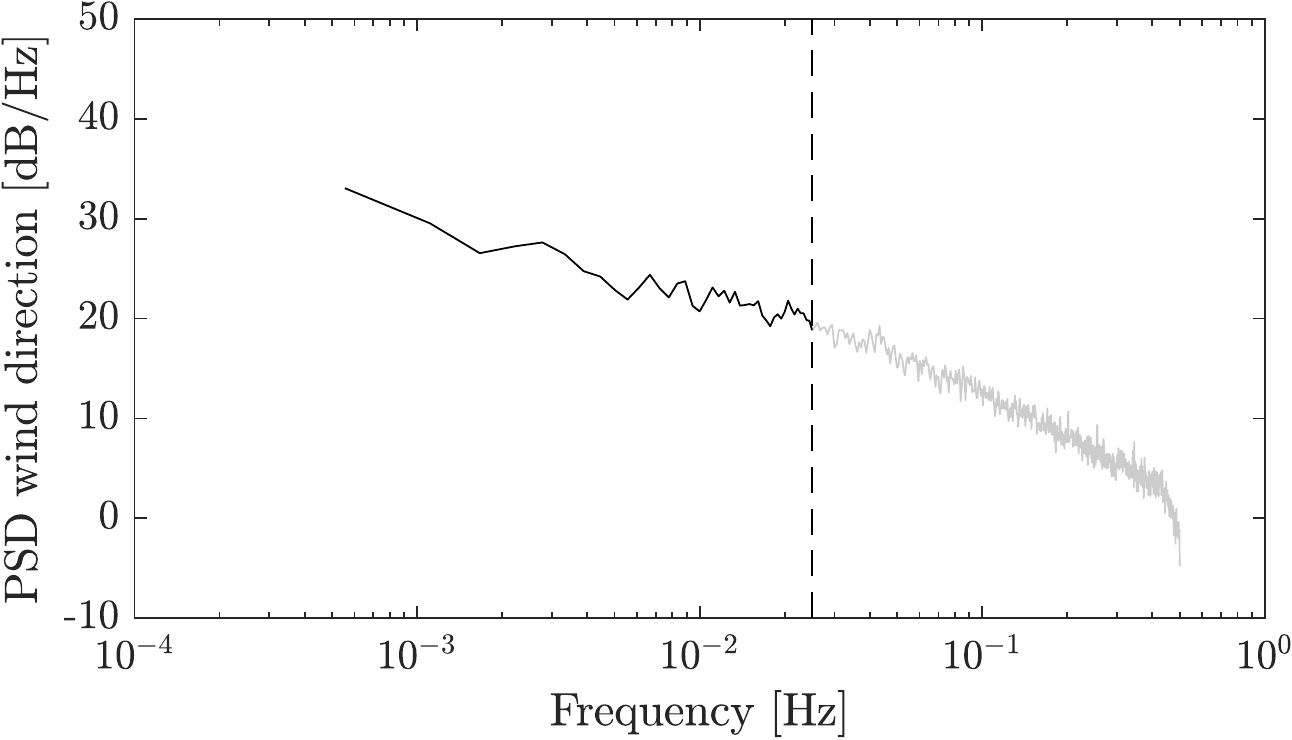}
    \caption{Power spectral density of the wind direction variation based on 1-Hz measurement data. \vspace{.5mm}}
    \label{fig:windSpectrum}
    \end{minipage}\hspace{0.02\textwidth}%
    \begin{minipage}{0.49\textwidth}
    \includegraphics[width=1\textwidth]{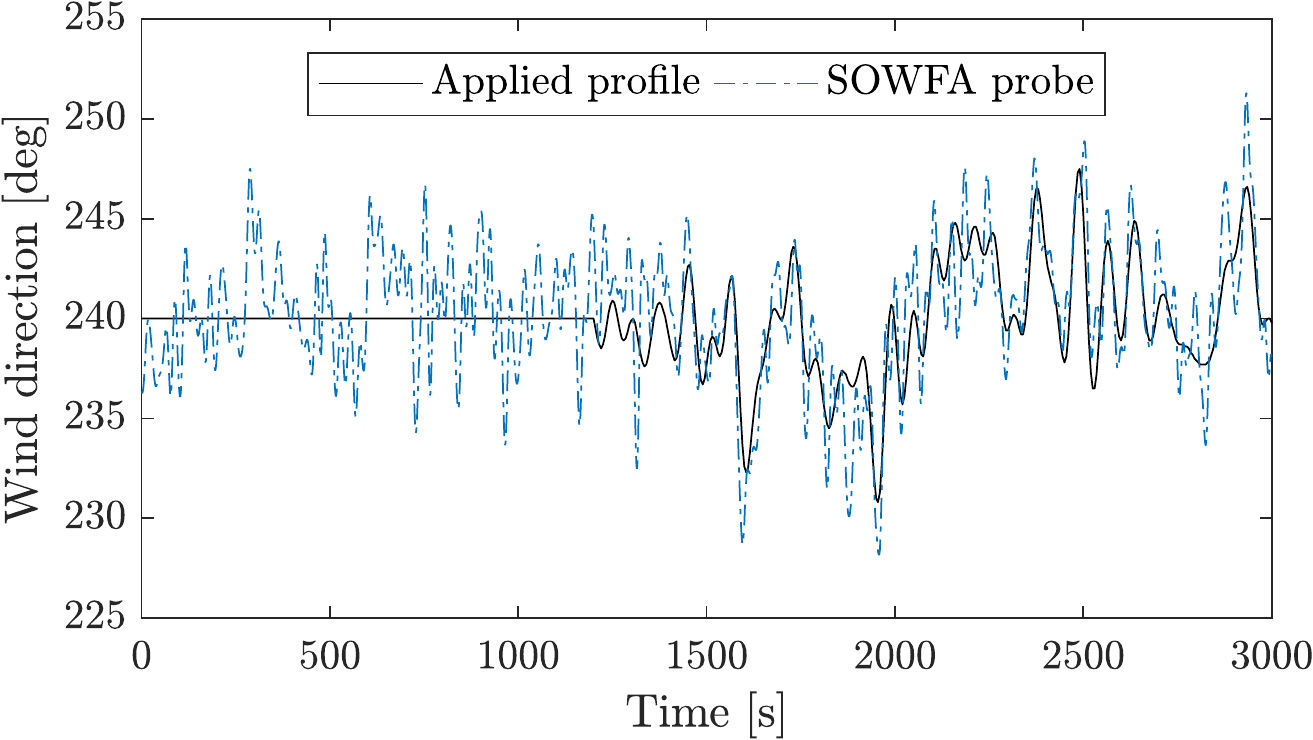}
    \caption{Wind direction profile as implemented in SOWFA (solid) compared to a point measurement of the flow field (dashed).}
    \label{fig:windProfile}
    \end{minipage} 
\end{figure}

In order to replicate realistic operating conditions at the wind farm site, the simulations are performed using an inflow wind field with a time-varying wind direction. This is achieved in SOWFA by adjusting the boundary pressure gradients. The wind direction profile that is used for the simulations is based on high-frequency measurements from an onshore met mast. Multiple thirty minute time-series of the measured wind direction were collected to compute the power spectral density given in \Cref{fig:windSpectrum}. This spectrum is subsequently used to generate a random wind direction profile based on the low-frequency content of the measurements, similar to \cite{Simley2019}. Furthermore, the profile is scaled to match the 3$^{\circ}$ standard deviation of the wind direction that was obtained from the met mast data near the Amalia wind farm site. High-frequency wind direction changes are not taken into account for this profile, as this is largely the result of local turbulence that is already present in the simulated wind field, as can be seen in \Cref{fig:windProfile}.  

\section{Framework for estimating the annual energy gain} \label{sec:methodology}
This section describes the overall framework for estimating the annual energy gain due to wake steering. First, a short description of Gaussian processes is given in \Cref{sec:GPopt}. \Cref{sec:yawOpt} discusses the optimization of the yaw angles. \Cref{sec:AEPfit} presents how the simulation results from FLORIS and SOWFA are integrated. 

\subsection{Gaussian processes} \label{sec:GPopt}
A Gaussian process can be described as a non-parametric model consisting of a mean and covariance function. It assumes that a set of function values belongs to a multivariate Gaussian distribution. The correlation between function values is determined by a covariance function based on the respective inputs and a set of hyperparameters. In this paper, the squared exponential covariance function is used. For a given set of training data, a Gaussian process is able to infer function values at new input locations. One of the strengths of Gaussian processes is their ability to estimate their own uncertainty, which inherently allows them to deal with uncertainty in the training data. This aspect is relevant for our AEP prediction framework, as we combine simulation results from models with different uncertainty levels. For a more detailed explanation of Gaussian processes, the reader is referred to \cite{rasmussen2006gaussian}. Gaussian processes can subsequently be used as a surrogate model for Bayesian optimization \cite{Shahriari2016}. This property will be used in \Cref{sec:yawOpt} to optimize the yaw angles of the wind farm.  

\subsection{Yaw angle optimization} \label{sec:yawOpt}
The wind turbine yaw angles are optimized over the entire wind rose for $1^{\circ}$ wind direction increments and below rated wind speeds ranging from 4 to 13 m/s. Simulation results from FLORIS are used to obtain a Gaussian process model of the wind farm power as a function of the yaw misalignments. This model is subsequently used to optimize the yaw misalignments. 

The search space of the yaw misalignments is limited to positive angles with a maximum of 30$^{\circ}$, for a number of reasons. Firstly, the search space is significantly reduced, thereby requiring less evaluations of FLORIS for convergence. Secondly, when optimizing the yaw angles for both positive and negative yaw misalignments, the optimal solutions might result in an absolute change of yaw orientation of over $50^{\circ}$ for a $1^{\circ}$ wind direction change (see for example \cite{Kanev2018}). Such a steep change in the yaw orientation could lead to a large increase in yaw activity. Additionally, it has been demonstrated in high-fidelity simulation studies and field experiments that negative yaw misalignments result in higher loads for yawed turbines compared to positive yaw \mbox{misalignments \cite{Fleming2014,Gebraad2016,Damiani2018}}. Finally, high-fidelity simulations have shown that negative yaw misalignments are less effective than positive misalignments for increasing the power output of a wind farm. This is thought to be a combined effect of the clockwise rotation of the turbine rotor (around the horizontal axis) and the Coriolis effect (in the northern hemisphere) \cite{Gebraad2016,ARCHER201934}.

The yaw misalignments are optimized robustly to account for the time-varying wind \mbox{direction \cite{Rott2018}}. 
This consists of optimizing over multiple wind directions using a weighted sum of the power signals $P_j$ with a probability density function $\rho(\phi)$:
\begin{equation}
    \centering
    \bm{\mathit\Gamma}^{opt}(\phi) = \arg \max_{\bm{\gamma}}\int_0^{2\pi}\rho(\phi)\sum_{j=1}^n P_j(\gamma_j,\phi)d\phi.
\end{equation}
In practice, the Gaussian distribution is discretized at five wind directions consisting of the mean direction, and $\pm\sigma$ and $\pm 2\sigma$ from the mean wind direction. In this paper, $\sigma=3^{\circ}$ as determined from the measurement data, is used for the optimization. While robust optimization will reduce the potential increase in energy for a single wind direction, it will make the yaw setpoints less sensitive to changes in wind direction.

\subsection{Improving the annual energy gain prediction} \label{sec:AEPfit}
\begin{figure}[h]
    \centering
    \includegraphics[clip,trim=0.5cm 11cm 2.5cm 1.5cm,width=1\textwidth]{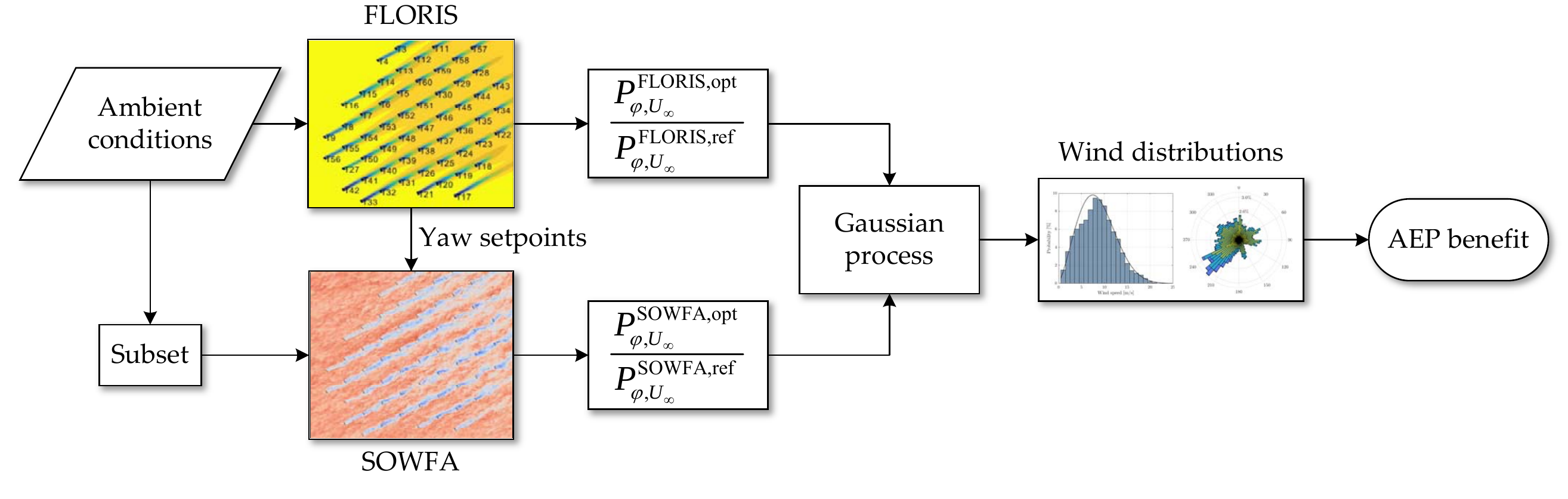}
    \caption{Flowchart for computing the improved AEP gain. \label{fig:AEPflowchart} }
\end{figure}

The entire framework for improving the prediction of the AEP benefit is given as a flowchart in \Cref{fig:AEPflowchart}. First, the surrogate wind farm model FLORIS is used to optimize the yaw angles for a range of ambient conditions. A subset of these ambient conditions is simulated in SOWFA with the previously obtained yaw setpoints. The resulting thirty minute time-series are divided into multiple five minute time-series in order to increase the amount of data points, as well as incorporate the variance of the power gain resulting from a varying wind direction in the Gaussian process. Next, the power gains of both the FLORIS and SOWFA simulations are computed and combined into a single dataset. A Gaussian process model of the power gain as a function of wind speed and wind direction is subsequently computed from the training data. This model is combined with the wind distributions of the wind farm site to estimate the AEP benefit. 

\section{Results} \label{sec:results}
This section presents the results of the AEP gain prediction framework applied to the Amalia wind farm. \Cref{sec:SOWFAresults} discusses simulations results from SOWFA. In \Cref{sec:AEPresults}, a Gaussian process model of the energy gain is obtained and the overall AEP gain is computed.

\subsection{High-fidelity simulations} \label{sec:SOWFAresults}
\begin{figure}[h]
    \centering
    \includegraphics[width=0.7\textwidth]{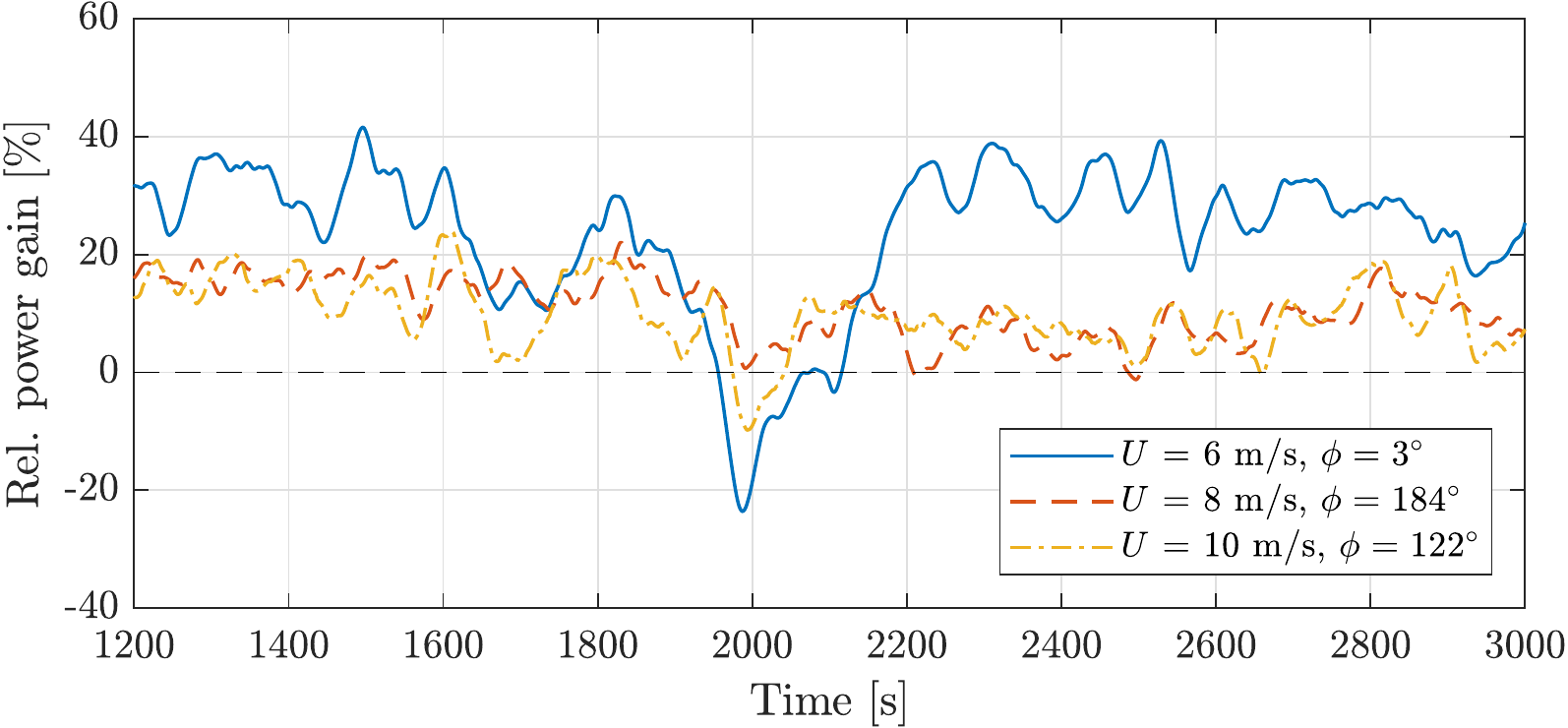}
    \caption{The relative power gain of the wind farm over time for different ambient conditions.}
    \label{fig:timeSeries}
\end{figure}

The Amalia wind farm is simulated in SOWFA for 50 different sets of ambient conditions. In \Cref{fig:timeSeries}, the relative power gains over time for different ambient conditions are given. The first 1200 seconds of the simulation are not considered in order to let the wakes propagate and remove any turbine start-up behaviour. It can be seen that especially for wind speeds of 6 m/s, wake steering shows larger gains than for higher wind speeds. This is believed to be partially the result of the turbine controller implementation of the NREL 5MW turbine, which operates the turbine at a sub-optimal tip-speed ratio for wind speeds near cut-in. As the wind speed increases, the turbine starts operating closer to the optimal tip-speed ratio, thereby increasing the efficiency of the turbine. In this case, wake steering not only reduces the wind speed deficit for waked turbines -- it also lets them operate more efficiently. 

\begin{figure}[h]
    \begin{minipage}{0.49\textwidth}
        \includegraphics[width=1\textwidth]{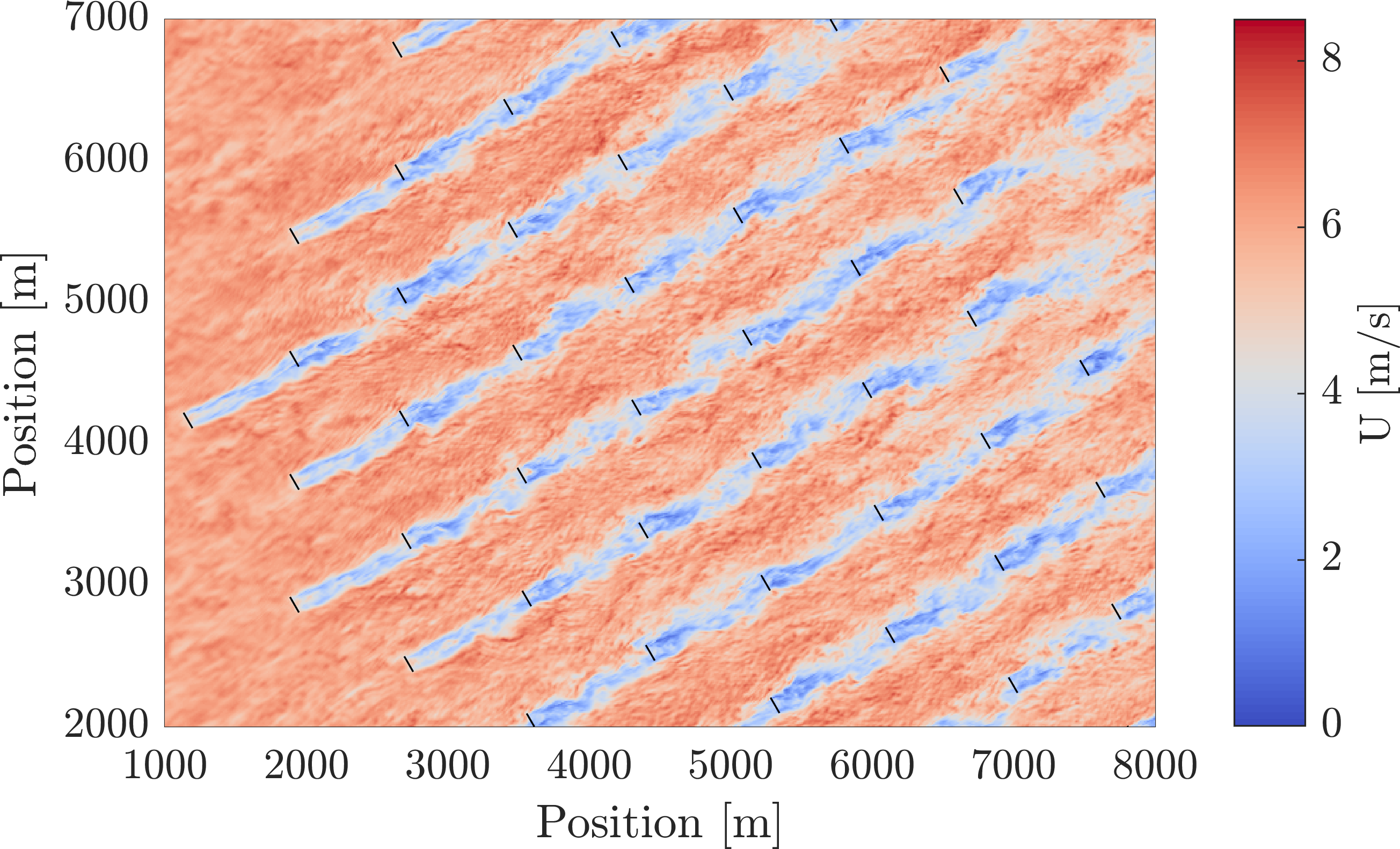}
    \end{minipage}\hspace{0.02\textwidth}%
    \begin{minipage}{0.49\textwidth}
        \includegraphics[width=1\textwidth]{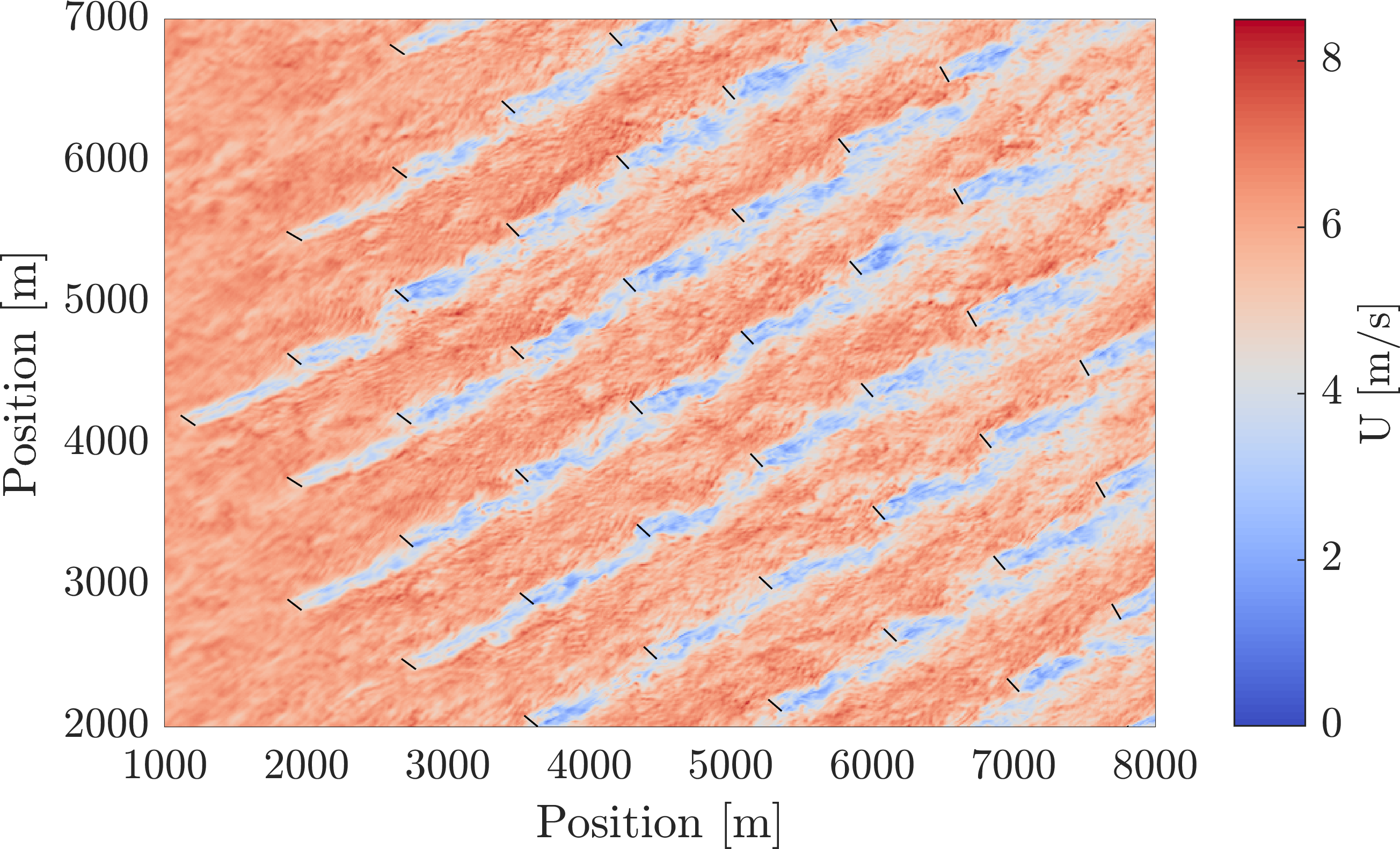}
        \end{minipage} 
    \caption{Velocity profile at turbine hub height of part of the wind farm with baseline operation (left) and wake steering (right) at time instant $t=2250$ s, for $U_{\infty} = 6$ m/s and $\phi_{avg}=3^{\circ}$.}
    \label{fig:SOWFAsim_1}
\end{figure}

\begin{figure}[H]
    \begin{minipage}{0.49\textwidth}
        \includegraphics[width=1\textwidth]{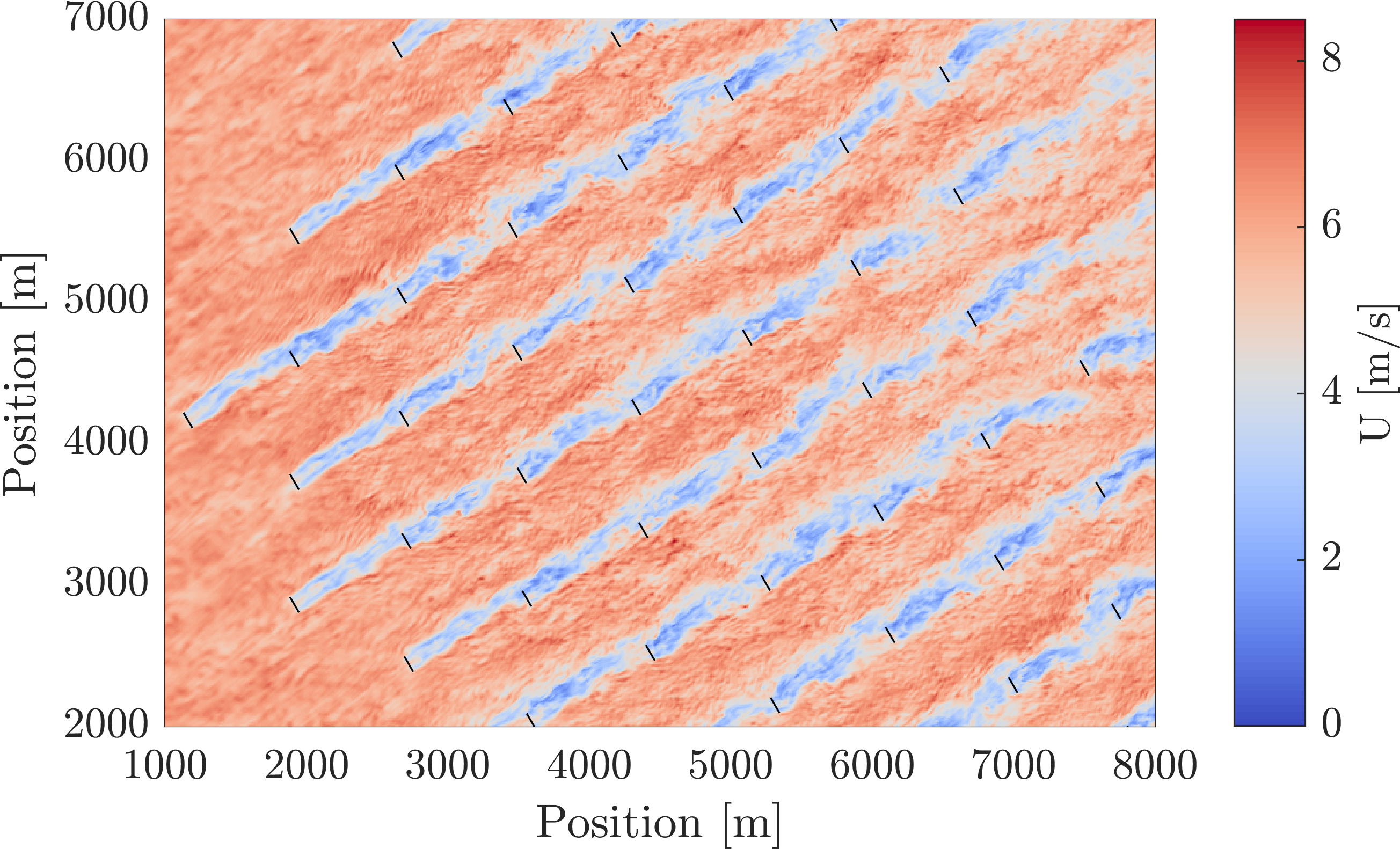}
    \end{minipage}\hspace{0.02\textwidth}%
    \begin{minipage}{0.49\textwidth}
        \includegraphics[width=1\textwidth]{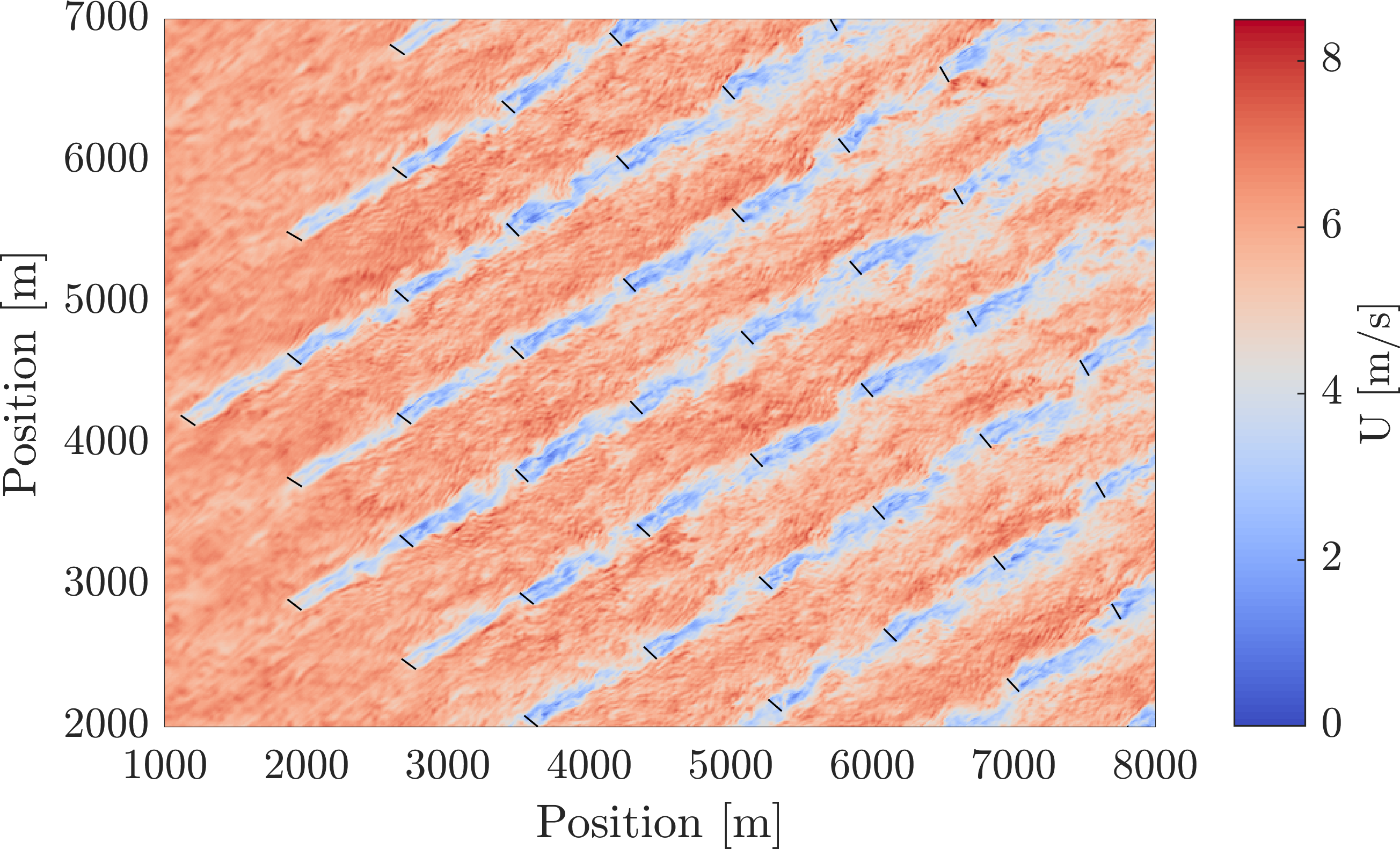}
        \end{minipage} 
    \caption{Velocity profile at turbine hub height of part of the wind farm with baseline operation (left) and wake steering (right) at time instant $t=1980$ s, for $U_{\infty} = 6$ m/s and $\phi_{avg}=3^{\circ}$.}
    \label{fig:SOWFAsim_2}
\end{figure}

Wake steering is able to achieve large gains in power at specific time instants. In \Cref{fig:SOWFAsim_1}, baseline operation is compared to wake steering for one of these instants. The figure clearly shows that wake steering is able to deflect the wake away from downstream turbines. Large negative gains are also observed for some periods of time in \Cref{fig:timeSeries}. Due to the time-varying wind direction, the wind farm operates with unfavorable yaw orientations at times. This is visualized in \Cref{fig:SOWFAsim_2}, where it can be seen that the yawed turbines redirect the wake into downstream turbines, resulting in a temporary decrease in power compared to baseline operation. Overall, the relative power gain remains significant, even in the presence of a time-varying wind direction.  

\subsection{Annual energy gain} \label{sec:AEPresults}
\begin{figure}[h]
    \begin{minipage}{0.44\textwidth}
        \includegraphics[width=1\textwidth]{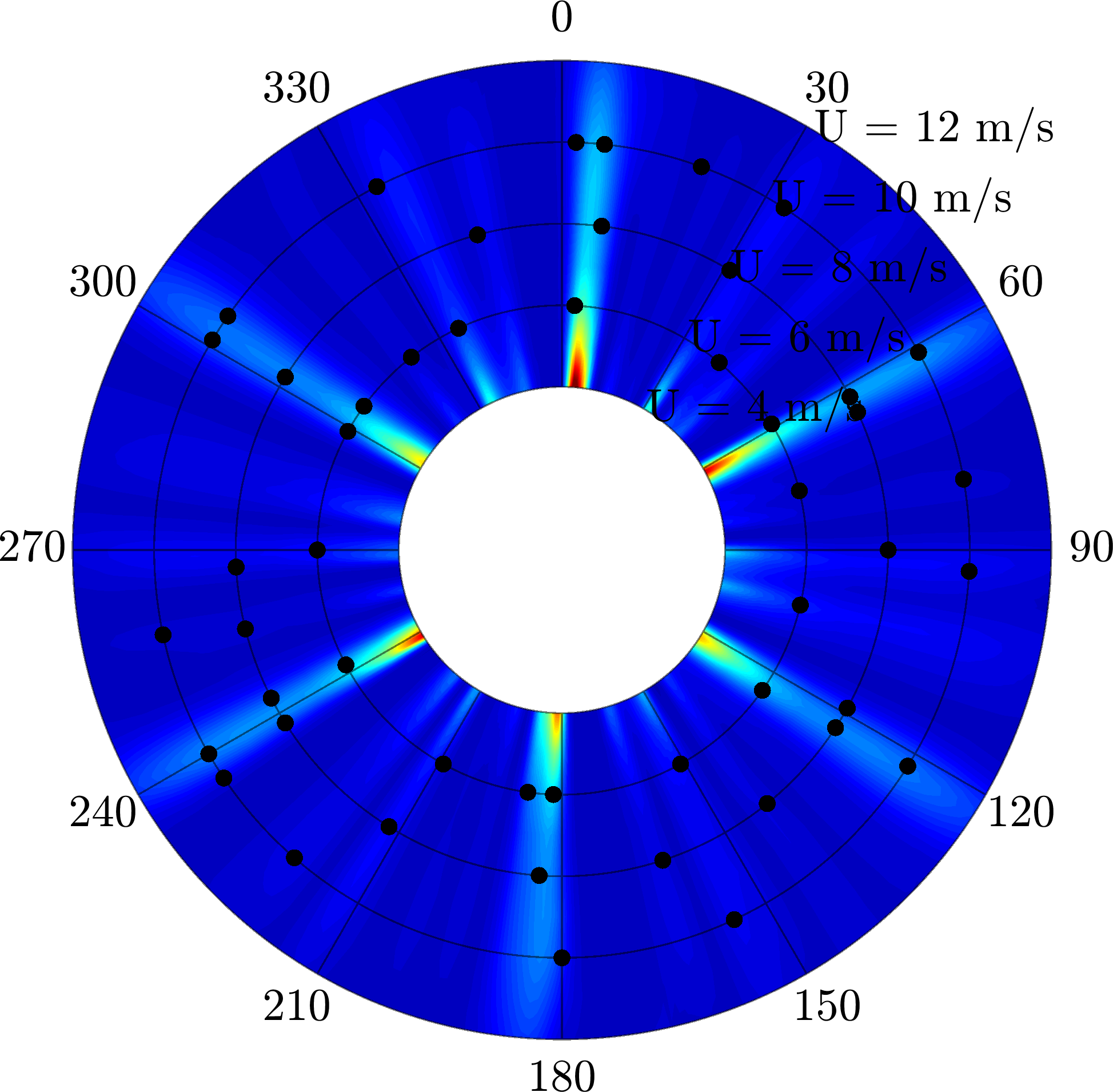}
        \caption{Robustly evaluated power gains according to FLORIS. The black dots represent the cases that are evaluated in SOWFA.}
        \label{fig:powerGainPolarFLORIS}
    \end{minipage}\hspace{0.02\textwidth}%
    \begin{minipage}{0.54\textwidth}
        \includegraphics[width=1\textwidth]{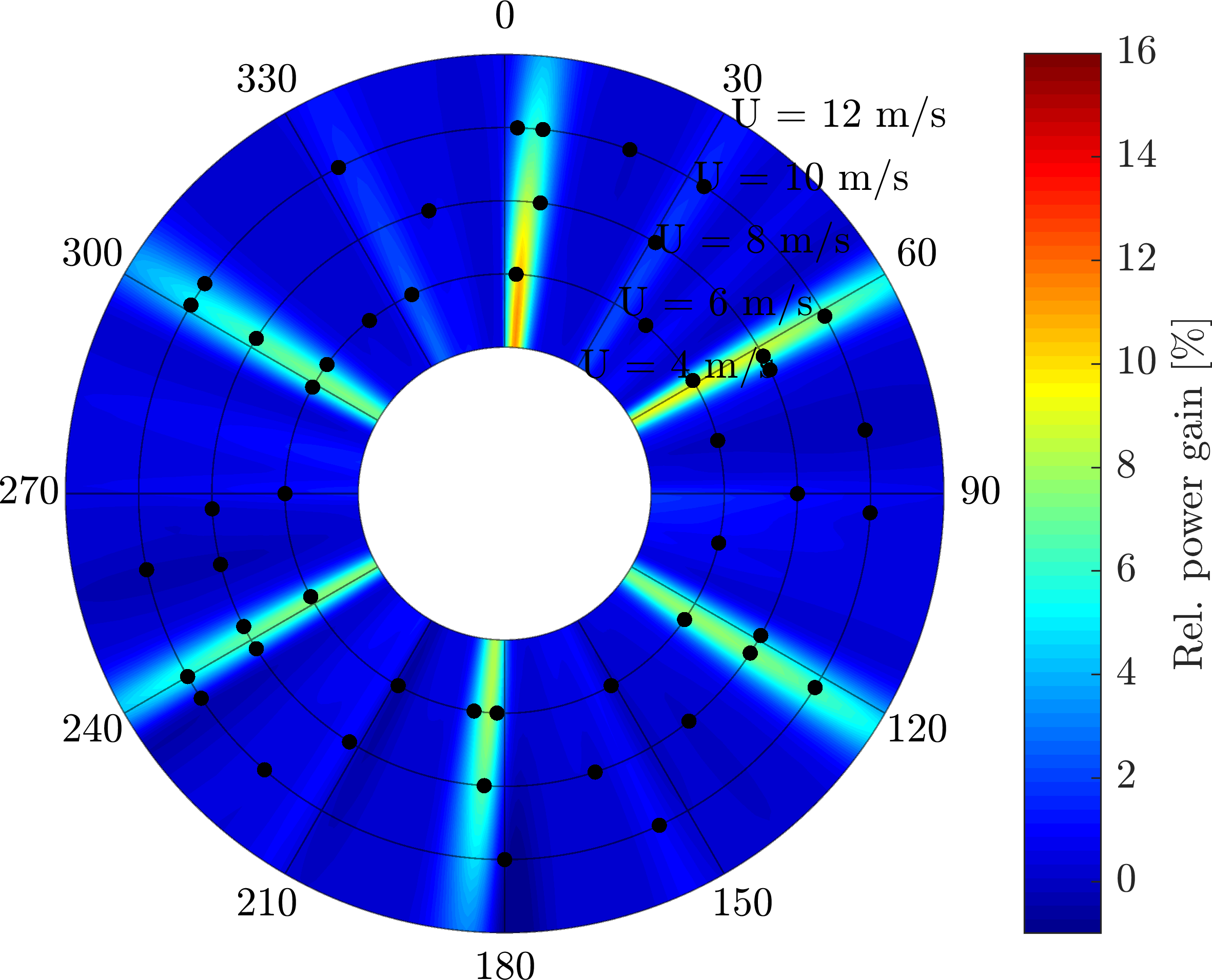}
        \caption{Power gains obtained through Gaussian process regression with large eddy simulation results. The black dots represent the cases that are evaluated in SOWFA.}
        \label{fig:powerGainPolarSOWFA}
        \vspace{1mm}
    \end{minipage} 
\end{figure}

In order to estimate the annual energy gain, the performance is evaluated robustly \cite{Rott2018}, i.e., the optimal yaw angles are evaluated for different wind directions using the approach from \Cref{sec:yawOpt}. The resulting power gains computed with FLORIS are presented in \Cref{fig:powerGainPolarFLORIS}. The steady-state simulation results show six principal wind directions, corresponding to a turbine spacing of 7D, where large gains in power are predicted. At an ambient wind speed of 4 m/s, a number of turbines are not generating any power under baseline operation. However, with wake steering the wind speed increases above cut-in, resulting in large gains.

Next, the combined results from FLORIS and SOWFA are used to obtain a model of the energy gain through Gaussian process regression. \Cref{fig:powerGainPolarSOWFA} presents the resulting Gaussian process model of the energy gain. The improved model shows increased gains for wind speeds of 6 m/s and above when compared to FLORIS. The Gaussian process model gives slightly smaller gains at lower wind speeds due to the regression with the SOWFA simulation data at 6 m/s.

\begin{figure}[h]
    \centering
    \includegraphics[width=0.7\textwidth]{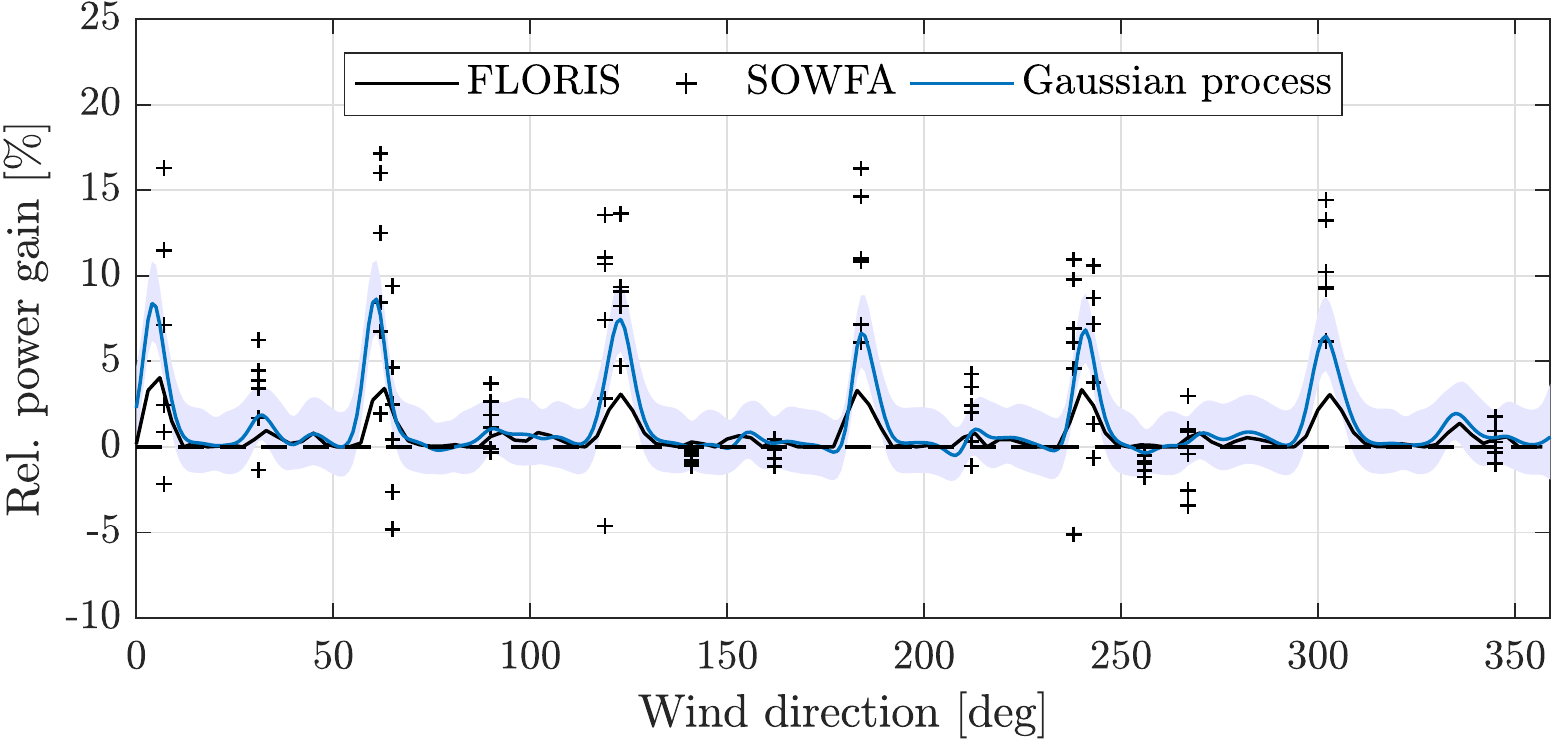}
    \caption{Relative power gain of the wind farm due to wake steering as a function of the wind direction, a wind speed of $8$ m/s and turbulence intensity of $5\%$. The light blue region indicates the 95\% confidence bounds of the Gaussian process regression model.
    }
    \label{fig:GPpowerGain}
\end{figure}

A more detailed comparison of the two models is provided in \Cref{fig:GPpowerGain}. This figure shows the energy gain as a function of wind direction for a single wind speed. For wind directions where FLORIS predicted a high gain in power, the Gaussian process model shows gains that are up to twice as high. This large difference in power between the two simulation models can possibly be ascribed to the secondary steering effects \cite{Martinez2018}, which are not modeled in our implementation of FLORIS. For the remaining wind directions, the simulation results from both simulation models are more in line. 

Finally, the energy gains predicted by both models are combined with the wind distribution data to provide an estimate of the AEP benefit with wake steering. The resulting gains are presented in \Cref{tab:AEP_results}. The AEP benefit according to FLORIS is smaller compared to previously estimated values for the Amalia wind farm, which were in the range of 1.10-1.28\%  \cite{Fleming2016journal,Kanev2018}. This is due to the fact that AEP was evaluated robustly to account for a varying wind direction. With the improved energy gain model, the AEP benefit shows a 76\% increase compared to FLORIS. This demonstrates the capability of the proposed framework for determining the AEP benefit. Furthermore, the difference between the two models emphasizes the need of further improvements to FLORIS, especially for large wind farms. Finally, it shows that wake steering can still be successful with a time-varying wind direction. In this paper, the yaw setpoints were fixed for the entire duration of the simulation. By updating the yaw angles as the wind direction changes, higher power gains could possibly be achieved. 

\begin{table}[h]
    \caption{AEP benefit with wake steering according to the two models.}
    \label{tab:AEP_results}
    \rm
    \centering
    \vspace{3mm}
    \begin{tabular}{@{}*{7}{l}}
        \br
        Model & AEP gain\\
        \mr
        FLORIS & 0.34\%\\
        Gaussian process & 0.60\%\\
        \br
    \end{tabular}
\end{table}

\section{Conclusion} \label{sec:conclusion}
In this paper, a novel framework was developed for determining the benefit of wake steering on the annual energy production of a large offshore wind farm. The framework uses a simplified surrogate model to optimize the yaw angles and provide an initial estimate of the energy gain. Next, large eddy simulations incorporating a time-varying wind direction profile were performed for a subset of ambient conditions, in order to obtain a more accurate evaluation of wake steering. Simulation results from both models were subsequently used to fit a model of the energy gain through Gaussian process regression. The improved model showed a 0.60\% gain in annual energy production. This is a 76\% increase compared to the estimate obtained with the surrogate model. It is expected that higher gains can be achieved when the yaw angles are updated as the wind direction changes. 

\bibliographystyle{iopart-num}
\bibliography{references}
\end{document}